

\input harvmac

\ifx\epsfbox\UnDeFiNeD\message{(NO epsf.tex, FIGURES WILL BE IGNORED)}
\def\figin#1{\vskip2in}
\else\message{(FIGURES WILL BE INCLUDED)}\def\figin#1{#1}\fi
\def\ifig#1#2#3{\xdef#1{fig.~\the\figno}
\goodbreak\midinsert\figin{\centerline{#3}}%
\bigskip\centerline{\vbox{\baselineskip12pt
\advance\hsize by -1truein\noindent\footnotefont{\bf Fig.~\the\figno:} #2}}
\bigskip\endinsert\global\advance\figno by1}

\overfullrule=0pt
\def\Title#1#2{\rightline{#1}\ifx\answ\bigans\nopagenumbers\pageno0\vskip1in
\else\pageno1\vskip.8in\fi \centerline{\titlefont #2}\vskip .5in}

\font\ticp=cmcsc10

\font\secfont=cmcsc10

%
%
\baselineskip=18pt plus 2pt minus 2pt

\def\ajou#1&#2(#3){\ \sl#1\bf#2\rm(19#3)}
%
\def\CH{{\cal H}}

\def\CM{{\cal M}}

\def\CZ{{\cal Z}}
\def\sN{\scriptscriptstyle N}

%


\def\z{ z}
%


\def\s{\sigma}

\def\t{\tau}
\def\a{\alpha}
\def\b{\beta}
\def\d{\delta}

\def\p{\pi}
\def\e{\epsilon}

\def\D{\Delta}

\def\zn{Z_N}
%


\def\TrH#1{ {\raise -.5em
                      \hbox{$\buildrel {\textstyle  {\rm Tr } }\over
{\scriptscriptstyle \CH _ {#1}}$}~}}
\def\char#1#2{\vartheta{\bigl[ {#1\atop  #2}\bigr] }}
\def\charbig#1#2{\vartheta{\biggl[ {#1\atop  #2}\biggr] }}
\def\det{Det {\bigl[ {a\atop b}\bigr] }}
\def\IZ{\relax\ifmmode\mathchoice
{\hbox{\cmss Z\kern-.4em Z}}{\hbox{\cmss Z\kern-.4em Z}}
{\lower.9pt\hbox{\cmsss Z\kern-.4em Z}}
{\lower1.2pt\hbox{\cmsss Z\kern-.4em Z}}\else{\cmss Z\kern-.4em Z}\fi}
\def\IC{\relax\hbox{$\inbar\kern-.3em{\rm C}$}}
\def\IR{\relax{\rm I\kern-.18em R}}
\def\1{\relax 1 { \rm \kern-.35em I}}
\font\cmss=cmss10 \font\cmsss=cmss10 at 7pt
\def\kN{\frac{k}{ \sN}}
\def\lN{\frac{l}{ \sN}}
\def\WPM{{{d^2\!\tau} \over {({\rm Im}\,\tau)^2}}}
\def\ad{{\dot a}}
\def\dN{\delta = 2\pi ( 1-{1\over N } )}
%

\def\frac#1#2{{#1 \over #2}}
\def\ie{{\it i.e.}}
\def\eg{{\it e.g.}}

\def\p+{{\partial_+}}

\def\half{{1 \over 2}}
\def\ket#1{|#1\rangle}
\def\pbar{{\bar p}}
\def\dbar{{\bar d}}
\def\mbar{{\bar m}}
\def\tb{{\bar \tau}}
\def\abar{{\bar \alpha}}
\def\zbar{{\bar z}}
\def\etab{\bar \eta}
\def\atil{{\tilde \a}}
\def\Stil{{\tilde S}}
\def\grad{\bigtriangledown}
\def\imt{{\rm Im}\tau}

%


%

\Title{\vbox{\baselineskip12pt\hbox{HUTP-94-A019}
\hbox{hep-th/9408098}
}}
{\vbox{\centerline {\bf  STRINGS ON A CONE AND BLACK HOLE ENTROPY}}}

\centerline{{\ticp Atish Dabholkar}\footnote{$^*$}
{Address after September 1, 1994: 452-48, Caltech,
Pasdena, CA 91007, USA \semi
\indent email: atish@theory.caltech.edu }}
\vskip.1in
\centerline{\it Lyman Laboratory of Physics}
\centerline{\it Harvard University}
\centerline{\it Cambridge, MA 02138-2901, USA}
\centerline{ e-mail: atish@string.harvard.edu}
\vskip .1in

\bigskip
\centerline{ABSTRACT}
\medskip

String propagation on a cone with deficit angle $2\pi (1- \frac{1}{N} ) $  is
described  by constructing a non-compact orbifold of a plane by a $Z_{N}$
subgroup of rotations.
It is modular invariant and has tachyons in the twisted sectors that
are localized to the tip of the cone.
A possible connection with the quantum corrections to the
black hole entropy is outlined. The entropy computed by
analytically continuing in N would receive contribution only from
the twisted sectors  and  be naturally proportional to
the area of the event horizon. Evidence is presented for
a new duality for these orbifolds similar to the
${\scriptstyle R} \rightarrow {1\over R} $ duality.

\bigskip

\bigskip

\vfill\eject

\def\np#1#2#3{{\sl Nucl. Phys.} {\bf B#1} (#2) #3}
\def\pl#1#2#3{{\sl Phys. Lett.} {\bf #1B} (#2) #3}
\def\prl#1#2#3{{\sl Phys. Rev. Lett. }{\bf #1} (#2) #3}
\def\prd#1#2#3{{\sl Phys. Rev. }{\bf D#1} (#2) #3}

\def\cmp#1#2#3{{\sl Comm. Math. Phys. }{\bf #1} (#2) #3}

\lref\thooft{G. 't Hooft,  \np{256}{1985}{727}.}
\lref\sussuglu{L. Susskind and J. Uglum,
{\it Black Hole Entropy in Canonical Quantum Gravity and  Superstring~Theory},
Stanford   Preprint SU-ITP-94-1 (1994),  hep-th/9401070.}
\lref\bombetal{L. Bombelli, R. Koul, J. Lee and R. Sorkin,  \prd{34} {1986}
{373}. }
\lref\callwilc{C. G. Callan and F. Wilczek, \pl{333}{1994}{55},
hep-th/9401072.}
\lref\kabastra{D. Kabat and M. J. Strassler, \pl{329}{1994}{46},
hep-th/9401125.}
\lref\dowker{J. S. Dowker,  {\sl Class. Quant. Gravity} {\bf 11} (1994) L55,
hep-th/9401159.}
\lref\srednick{M. Srednicki, \prl{71}{1993}{666}.}
\lref\bekenste{J.  D. Bekenstein,  \prd{7}{1973}{2333}; \prd{9}{1974}{3292}. }
\lref\hawking{S. W. Hawking, \prd{14}{1976}{2460};
\cmp{43}{1975}{199}; \cmp{87}{1982}{395}.}
\lref\gibbhawk{G. W. Gibbons and S. W. Hawking,
\prd{15}{1977}{2752}\semi
S. W. Hawking, \prd{18}{1978}{1747}.}
\lref\birrdavi{N. D. Birrell and P. C. Davies, {\it Quantum Fields in Curved
Space}, Cambridge
University Press (1982).}
\lref\inprep{A. Dabholkar, {\it Quantum Corrections to Black Hole
Entropy in String Theory}, Caltech Preprint CALT-68-1953 (1994),
hep-th/9409158}
\lref\teitelbo{M. Ba{$\tilde {\rm n}$}ados, C. Teitelboim  and J. Zanelli,
\prl{72}{1994}{957}, hep-th/9309026.}
\lref\carlteit{S. Carlip, C. Teitelboim, {\it The off Shell Black Hole}, IAS
preprint IASSNS-HEP-93-84 (1993), gr-qc/9312002.}
\lref\dixoetal{L. Dixon, J. Harvey, C. Vafa and E. Witten,
\np{261}{1985}{678};  \np{274}{1986}{285}. }
\lref\grosetal{D. J. Gross, M. J. Perry and L. G. Yaffe, \prd{25}{1982}{330}.}
\lref\dewitt{B. S. DeWitt, \prd{160}{1967}{1113}.}
\lref\hawktwo{S. W. Hawking, in {\it Recent Developments in Gravitation} ,
edited by  M. Levy, Plenum (1979). }
\lref\baggetal{ J. A. Bagger, C. G. Callan and  J. A. Harvey,  {\sl Nucl.
Phys.} {\bf B278} (1986) 550.}
\lref\stringpert{E. D'Hoker and D. H. Phong, {\sl Reviews of Modern Physics}
{\bf 60} (1988);
and references therein.}
\lref\alvarez{O. Alvarez, \np{276}{1983}{125}.}
\lref\moornels{G. Moore and P. Nelson, \np{266} {1986}{58}.}
\lref\susskind{L. Susskind, {\it Some Speculations about Black Hole Entropy
in String Theory}, Rutgers University preprint RU-93-44 (1993),
hep-th/9309145.}
\lref\seibwitt{ N. Seiberg and E. Witten, \np{276}{1986}{272}.}
\lref\fioletal{T. M. Fiola, J. Preskill, A. Strominger, S. P. Trivedi,
{\sl Black Hole Thermodynamics and Information loss in Two Dimensions}, Caltech
Report
CALT-68-1918 (1994), hep-th/9403137.}
\lref\alvaetal{L. Alvarez-Gaum{\' e}, G. Moore and C.  Vafa,
\cmp{106} {1986}{40}.}
\lref\greenbook{M. B. Green, J. H. Schwarz and E. Witten,
{\it Superstring Theory},  {\rm vols. I and II} , Cambridge University Press
(1987).}
\lref\mumford{D. Mumford, {\it Tata Lectures on Theta {\rm I}}, Birkh{\"a}user
 (1983).}
\lref\tseytlin{A. A. Tseytlin, {\sl International Journal of Modern Physics}
{\bf A4}, No. 6 (1989)
1257. }
\lref\ginsparg{P. Ginsparg, {\sl Applied Conformal Field Theory}, Les Houches
Session XLIV,
1988 in {\it Fields, Strings and Critical Phenomena} ed. by  E.  Br\'ezin and
J. Zinn-Justin,
North Holland (1989).}
\lref\bowgidd{M. J. Bowick and S. B. Giddings, \np{325}{1989}{631}.}
\lref\aticwitt{J. Atick and E. Witten, \np{310}{1988}{291}.}
\lref\satkog{B. Sathiapalan, \prd{35}{1987}{3277}\semi
Ya. I. Kogan {\sl JETP Lett.} {\bf 45} (1987) 709.}
\lref\susstwo{L. Susskind, \prd{49}{1994}{6606};
\prl{71}{1993}{2367}.}

\newsec{Introduction}

In this paper we discuss the propagation of strings on a conical space.
The chief motivation for this work
stems  from its possible application to computing the entropy of a black
hole in string theory.
In field theory, an efficient way to compute the entropy  is via the
Euclidean path integral on a cone.
The leading contribution to the entropy comes from
field modes near the horizon of the black hole which in Euclidean space
corresponds to the tip of a cone.
In order to understand this leading contribution,
the details of the geometry of a specific black  hole can be ignored
and it is adequate to consider field propagation in a
conical space.
We would like to do something similar in string theory.
This work  should be regarded as  a  step in that direction.
For special values of the deficit  angle of the cone,
it is easy to construct the corresponding string theory
as a $Z_N$ orbifold of the theory in flat space. We describe
this construction in the next section.
These orbifolds can also be viewed
as describing  string propagation
in the background of a cosmic string.
We shall also see some evidence for  duality
in a new guise.
Moreover, for large N, there are nearly massless tachyons in the
spectrum that are  localized to the
tip of the cone.
These could be useful for studying the tachyon instability in string theory.

We shall  now  describe some  basic aspects of   black-hole entropy
that will be  important  in the following discussion.
There are three distinct notions of the entropy associated with a black hole.
The Bekenstein-Hawking entropy is calculated  using thermodynamics and
quantum field theory in a fixed,
classical background of a black hole \refs{\bekenste , \hawking}.
It is given by
\eqn\Sbh{S_{BH} = {A \over {4 G \hbar}}}
where A is the area of the event horizon of the black hole and
$G$ is the renormalized Newton's constant.
The Gibbons-Hawking entropy \gibbhawk , on the other hand,  is obtained
by evaluating the full functional
integral of quantum gravity
around a saddle point which represents the Euclidean continuation
of the Schwarzschild solution.
Remarkably, it also gives the same expression for the entropy
as the Bekenstein-Hawking
formula with the renormalized Newton's constant replaced by its bare value.
It is natural to ask about the quantum corrections to the leading
semiclassical formula.
Susskind and Uglum \sussuglu\ have argued that
these quantum corrections
can all be absorbed into the renormalization of Newton's constant.
Thus the renormalized Gibbons-Hawking entropy equals the Bekenstein-Hawking
entropy.

Both these derivations do not offer any statistical interpretation of
the thermodynamic black hole entropy in terms of counting of states.
't Hooft \thooft\ has advocated that the entropy of quantum fluctuations
as seen by a Schwarzschild observer should account
for the black hole entropy.
In field theory this quantity is ultraviolet divergent.
Several authors  have found a similar divergence in Rindler spacetime
which approximates the geometry of a large black hole very near the horizon
\refs{\sussuglu ,\srednick, \dowker , \bombetal , \callwilc , \kabastra}.
There is a simple description of the leading divergence.
A fiducial observer that is stationed at a fixed radial distance from the
black hole, has to accelerate with respect to the freely
falling observer in order not to fall into the black hole.
Very near the horizon, the fiducial observer is like a Rindler observer
in flat Minkowski space.
As a result, she sees a thermal bath \birrdavi\ at a
position-dependent proper temperature  ${ T(\z )} = {1\over{2\pi \z}}$ where
$z$ is the proper distance from the horizon.
Using Planck's formula for a single massless boson
we get the entropy density:
\eqn\fourentropy{
s(z) = {4\over 3} {\pi^2 \over 30} ({1\over{2\pi \z}})^3\,  .}
Note that we have been able to define the entropy density because
entropy is an {\it extensive} quantity as it should be.
However,  the dominant contribution comes from the region
near the horizon $z=0$ and is not extensive but proportional
to the area.   If  we put a cutoff on the proper distance  at
$z=\e$ (or alternatively on proper temperature) the total entropy is:
\eqn\totalentropy{\eqalign{
S &= \int _\e^\infty { s(z) A dz}\cr
   &=  {A \over {360\pi \e^2}}  \quad ,\cr}
}
where $A$ is the area in the transverse dimensions.
This is in agreement with the result obtained in
\refs{\thooft , \sussuglu}\  by other means.
Because the thermal bath is obtained by tracing over states that
are not accessible to the observer in the  Rindler wedge,
this entropy is  also the same as the `entropy of
entanglement'  \refs{\bombetal , \srednick , \kabastra ,  \fioletal }
or the `geometric entropy'  \refs{\dowker , \callwilc }.
For a massive field with mass $m$,
there will be corrections to this formula which will be down
by powers of $m\e$.

It is not clear how this divergent quantity can equal the black hole
entropy which is finite. There are other difficulties with
identifying this entropy of entanglement
with the Bekenstein-Hawking entropy.
For expample, the entropy of entanglement has no classsical
contribution and starts at one loop whereas the black hole entropy
is inversely proportional to the coupling constant.
Furthermore, the entropy of entanglement depends on the species
and couplings of various particles in the theory whereas the
black hole entropy does not.
't Hooft has argued that it is necessary to understand the
ultraviolet structure of the theory in order to address these issues.
He has conjectured that these difficulties will be resolved
once the correct short distance structure is known.
He has further suggested that this divergence of entropy in
field theory is intimately
related to the puzzle of loss of information in black hole evaporation.
If the entropy does have a statistical interpretation in terms of
counting of states then its divergence would suggest an infinite
number of states associated with  a finite mass black hole.
As long as the black hole has an event horizon, it can apparently store
an arbitrary amount of information in terms of correlations
between the outgoing radiation and the high energy modes near
the horizon. When the horizon eventually disappears,
the information in these correlations is irretrievably lost.

It is almost impossible to test these ideas within field theory,
especially when one is dealing with a nonrenormalizable theory
such as quantum gravity.
Fortunately, string theory offers a suitable
framework for addressing this question.
It is a perturbatively finite theory of quantum gravity
and comes with a well-defined matter content.
Moreover, Susskind \refs{\susstwo, \susskind}
has argued that string theory
may also possess some of the properties required for describing
black hole evaporation without information loss.
It is therefore of great interest to know how
the ultraviolet behavior of the entropy is controlled
in string theory.
One can hope that string theory
will illuminate this question in important ways.

In field theory, there is a general method of computing the entropy
of entanglement using Euclidean path integral over a conical space.
As we shall see in section three,  the analogous formula
in string theory for the entropy at $G$ loop is
\eqn\entropy{
S_{G}  \, = \, \frac{d (N \, A_{G} ) }{dN} |_{N=1} \, .
}
where $A_G$ is the vacuum amplitude  at $G$ loop
in string perturbation theory on a conical space with
deficit angle $\dN$.
A major obstacle in  using this formula is that
the conical background in general
does not satisfy the equations of motion
because there is a curvature singularity at the tip of the cone.
Luckily, in string theory, for special values of the deficit angle
$\delta = 2\pi (1- \frac{1}{N})$ with integer $N$,
the theory manages to be on-shell at least at tree level.
We would like to use this fact to obtain the entropy by analytically
continuing in $N$. Similar suggestion has been made also in  \callwilc .
We  do not yet have a complete expression for the
entropy and we wish to return to it in a subsequent publication \inprep .
However, the entropy computed this way already appears to have
several desirable features.
Moreover, a number of technical issues arise in the
construction of the orbifold
that are interesting in their own right.
With this objective in mind, in the next section, we
describe the propagation of strings
on a cone for integer $N$. In section three,  we discuss its
relation to the computation of entropy.

\newsec{ Strings on a Cone}

We first construct the bosonic orbifold from  the
uncompactified string theory in twenty-six dimensions. Our spacetime will be
$M_{24} \times K_{N}$ where $M_{24}$ is flat spacetime and $K_{N}$ is a cone
with deficit angle $2\pi (1-{1\over N})$.
We can further compactify some of the dimensions of the $M_{24}$ if
we so desire. The details of compactification will  not be important.
For non-integer N, this background is not a solution to the string equations
of motion because when a string encounters the  curvature singularity
at the tip of the cone, it would develop a kink.
For integer N, however,  we can have consistent
propagation of strings despite the curvature singularity.
\ifig\fone{Configurations on the plane that define consistent string
configurations on
the cone with deficit angle
${4\pi \over 3}$. The solid and the dashed lines indicate strings with zero and
nonzero winding
number around the tip of the cone respectively.}
{\epsfysize=2.0in \epsfbox{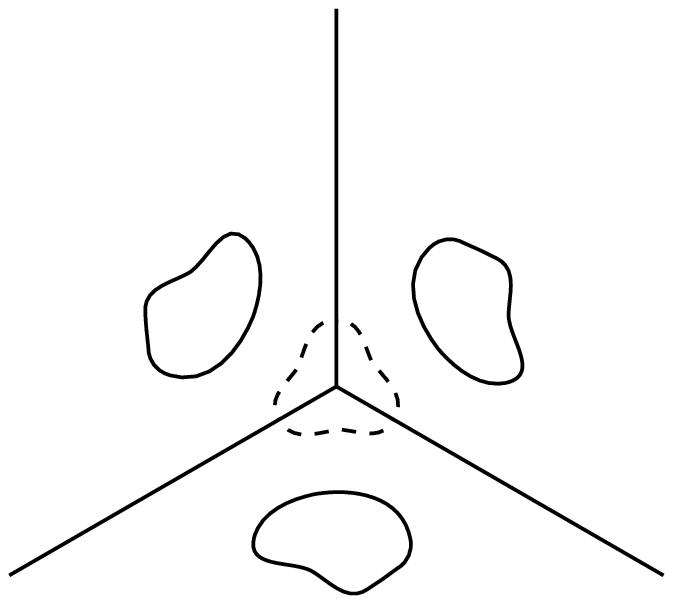}}
In this case,  we can
tile the entire plane with
N copies of the cone.
The configurations on
the plane that are symmetric under $Z_{N}$ rotations
define consistent string configurations on the cone
[see \fone ].
We can then regard $K_{N}$ as an orbifold of
the plane $R_{2}/{Z_{N}}$ \refs{\dixoetal , \ginsparg}.
Notice that unlike the orbifolds considered in string compactification
we are  interested in an orbifold of a non-compact space.
If instead of $R_2$ we consider a compact space like a torus \baggetal ,
then we cannot take $N$ to be arbitrary and the allowed
orbifold groups are very limited.  Moreover, in that case, there are
more than one points with a conical singularity and
the orbifold only locally looks like a cone.

The Hilbert space of the orbifold is obtained by first considering
the theory on the plane which is conformal and modular invariant and
then projecting  onto $\zn $ invariant states.  As is well known,  we also
have to include  the twisted string states because the winding number
on the plane
around the tip of the cone  is conserved only modulo $N$ [\fone ].
It is convenient to combine the co-ordinates of the plane into a complex boson
$X = \frac{X_1 +iX_2}{\sqrt 2}$ and ${\bar X} = \frac{X_1 - iX_2}{\sqrt 2}$.
The orbifold group then acts on $X$ by multiplication
by a phase $e^{2\pi i k \over N}$.
In the untwisted sector this field has the standard mode expansion:
\eqn\untwist{
X = x + p \t + \frac{i}{2} \sum_{n\not= 0} {\a_n \over n} e^{-2i n (\t - \s)}
 + \frac{i}{2} \sum_{n\not= 0} {\atil_n \over n} e^{-2i n (\t + \s)} .}
and the ground states are
labelled by the momenta in the plane $\ket{p , \pbar}$. The spectrum before
projection
is the same as the twenty-six dimensional string and
consists of states with some number of creation operators acting on the ground
states.
For states with nonzero $p$ and $\pbar$,
the projection onto  $\zn$ invariant  states reduces the spectrum
by a factor of $N$. For example, for $N=3$, the states
$\a_{-n}\ket{p , \pbar}$ are projected onto
$\frac{1}{3} (\a_{-n}\ket{p , \,\pbar} + e^{ \frac{2\pi i}{3}}  \a_{-n}\ket{
e^{-\frac{2\pi i}{3}} p , \,e^{ \frac{2\pi i}{3}} \pbar} + e^{ \frac{4 \pi
i}{3}}  \a_{-n}\ket{e^{ -\frac{4\pi i}{3}} p ,\, e^{ \frac{4\pi i}{3}}
\pbar})$.
Thus,  on a $\zn$ cone, we still have the same  set of particles in the
untwisted sector
as in flat space,  except that the
allowed combinations of momenta are reduced N-fold.
The zero momentum states, however,  need to be treated differently.
In this case, only those combinations
of   creation operators that are invariant under $\zn$ rotations are allowed.
For example,   $\a_{-n}\ket{0 , 0}$ is projected out but $\a_{-n} {\bar{ \atil}
}_{-m }\ket{0 , 0}$
is allowed. As we shall see,  it is the zero momentum sector that mixes with
the twisted sectors under modular transformations.
In the  twisted sectors, the boson is subject to  the boundary condition
$X(\s + \pi , \t ) = e^{2\pi i k \over N} X(\s, \t ) $, {k=0,\dots , N-1}.
The mode expansion and the commutation
relations are given by:
\eqn\twist{\eqalign{
&X =  \frac{i}{2} \sum_{n} {\a_{n +\kN}\over {n +\kN}} e^{-2i ({n+\kN}) (\t -
\s)}
 + \frac{i}{2} \sum_{n} {\atil_{n -\kN} \over {n -\kN}} e^{-2i({n -\kN} )(\t +
\s)} \, ,\cr
&   [\a_{m +\kN}, \abar_{n -\kN} ] =   (m+\kN ) \d_{m+n} \, ,\quad\quad
[\atil_{m +\kN}, \bar{\atil}_{n -\kN} ] = (m+\kN )\d_{m+n} \, .\cr
}}
The zero point energy for the  $\kN$  moded (complex) boson
is given by $\half \kN(1-\kN) -\frac{1}{12}$.  The $22$ untwisted bosons
contribute $\frac{-11}{12}$ . Consequently, we have
$N-1$ tachyons, one in each  twisted sector with mass-squared
$\frac{4k}{N}(1-\kN) -8$ .
Note that there are no zero modes in the expansion \twist .
As a result,
these tachyons are localized to the tip of the cone but can have
momenta in the remaining $24$ dimensions.

The vacuum amplitude  for the cone is not very different from the one in flat
space.
For the bosonic string in flat space,  the G-loop amplitude is given by
(see  \alvarez\ \moornels\ \stringpert\ )
\eqn\amplitude{
A _G \sim  g^{2G-2}  {\rm L}^{26} \int_{{\cal M}_G}
           {d(WP)\over {\rm vol(ker} P_1)}  \ \left({2\pi ^2 {\rm det}'\Delta
_g}\over
            {\int d^2\! z \sqrt g}\right) ^{-13} {\rm det}'(P_1^\dagger P_1)
}
Here $d(WP)$ is the Weil-Petersson measure over the genus-$G$ moduli space
${\cal M}_G$,
$\Delta _g$ is the scalar Laplacian $-\frac{1}{\sqrt g} \partial_a{\sqrt g}
g^{ab}\partial_b$ and
$P_1$ is the operator that maps vectors into symmetric traceless two-tensors:
$(P_1 v)_{ab} = \grad_a \, v_b +\grad_b \, v_a - g_{ab} \grad\, v$  .
The volume factor $ {\rm L}^{26}$ comes from the zero modes of the scalar
Laplacian, one
power for each  real boson.
${\rm det}^\prime$ is the determinant only over nonzero modes.
In the orbifold theory, one inverse power of the determinant in \amplitude\
coming from one complex free boson and  two powers
of L coming from the zero modes,
get replaced by the orbifold partition function $\CZ (N)$ for the field $X$, at
genus $G$.

It is easy to  write down the one-loop partition function explicitly.
The world-sheet at one loop is a torus.   The metric for a torus in a
given conformal class
is parametrized by a complex modular parameter $\t$: $ds^2 =|d\s_1 + \t
d\s_2|^2$
with $0 \leq \s_1 \, , \s_2 <1 $.
The partition function is given by the orbifold sum
\eqn\partition{
\CZ  (N) =  \sum_{k, l} \CZ_{k,l} (N) \ .
}
Each term  $\CZ _{k,l}$ represents the path integral of a complex  boson
on a torus with twisted  boundary conditions. The path integral gives one
inverse power of
a determinant $\det (\D_g )$ of the scalar Laplacian on a torus subject to the
boundary
conditions
\eqn\boundary{
X(\s_1 +1 \, , \, \s_2)  = e^{\frac{2\pi i k } {N} } \, X(\s_1 \, , \, \s_2)\,
,
 \quad X(\s_1 \, , \, \s_2+1 ) = e^{\frac{2\pi i l } {N} }\, X(\s_1 \, , \,
\s_2)  \, .
}
Instead of evaluating the bosonic determinant we shall evaluate a related
quantity,  $\det (\grad^z_{-\half})$
which is the determinant of a chiral Dirac
operator.
This will also be useful later when we discuss the superstring.
It can be regarded as a path integral of a complex chiral fermion with
boundary condition
$\psi (\s_1 +1 \, , \, \s_2) = -e^{2\pi i a} \psi (\s_1 \, , \, \s_2)$,
and $\psi (\s_1  \, , \, \s_2+1) = -e^{2\pi i b} \psi (\s_1 \, , \, \s_2) $.
It is straightforward  to evaluate this determinant in the operator
formalism \alvaetal .
Writing $q= e^{2\pi i \t}$,  and using the standard  relation between the path
integral and the operator formalism, it  is equal to the trace
$\TrH{}(h_b  \, q^ {H_a} )$.  $H_a$
is the Hamiltonian of a chiral, twisted fermion:
\eqn\hamiltonian{
H_a = \sum_{n=1} ^{\infty}(n-\half +a ) d_n^\dagger d_n +  (n-\half -a )
\dbar_n^\dagger \dbar_n
+\frac{a^2}{2} -\frac{1}{24}
}
The fermionic oscillators satisfy the canonical anticommutation relations
 $\{ d_n^\dagger , d_m\} = \d_{mn}$ and $\{ \dbar_n^\dagger , \dbar_m\} =
\d_{mn}$, and
 $\CH $ is the usual Fock space representation of these commutations.
The group $Z_N$ acts on this Fock space through
$ h dh^{-1} = -e^{-2\pi i b} d$ , $ h \dbar h^{-1} = -e^{2\pi i b} \dbar$.
The trace equals (up to an arbitrary phase)
\eqn\trace{
e^{2\pi i a b} q^{\frac{a^2}{2}-\frac{1}{24}}
\prod_{n=1}^{\infty}  (1+q^{n-\half +a} e^{2\pi i b} )\,
 (1+q^{n-\half - a} e^{-2\pi i b}) \ .
}
Using the  product representation  of the theta function $\char{a}{b}(\t)$
with characteristics \mumford\  ,   we see that
\eqn\detab{
\det (\grad^z_{-\half}) = \TrH{}(h_b  \, q^ {H_a} )
= \frac{\char{a}{b} (0|\t )} {\eta (\t )} \ ,
}
where  $\eta (\t )$ is the Dedekind $\eta  $ function.

Now we return to the problem at hand.
Up to zero modes, the chiral boson determinant is
the inverse of the chiral fermion determinant.
The orbifold sum  \partition\  becomes
\eqn\zN{
 \CZ (N)  =  \sum_{k,l=0}^{N-1} \Bigg| \frac
 {\eta (\t )} {\vartheta \big[ { {\frac{k} {\sN} +\half}  \atop {{\lN} +\half}
}\big] }
\Bigg| ^{2}
}
The bosonic zero modes can give  a divergence for some of these terms such as
$\CZ_{0,0}$,
and should be treated more carefully.
For the moment,  we shall continue to treat them in
a somewhat cavalier manner.

The Weil-Petersson measure  for the torus is $\WPM $ and
$ {\rm vol(ker} P_1)  = \imt $.
Using  the standard expressions for  the other determinants
\refs{\stringpert \greenbook}\ we obtain
\eqn\torusamp{
A_1 (N) \sim {\rm Area} \int_{\CM_1}
     {\WPM}   \quad  | \eta^2 (\t  ) \etab^2 (\tb ) \imt | ^{-11}  \quad
\frac{1}{N}\, \CZ (N)
}
The factor of $1\over N$ in this formula comes from the fact that the operator
that projects onto $Z_N$ invariant states is ${1\over N} (1 + h + h^2 + \dots +
h^{N-1}) $.
The area here refers to the volume $L^{24}$ of the transverse dimensions.
The integration is over the genus-one moduli space $ {\CM_1}$ which is
the fundamental domain of the modular group
\eqn\funda{
|\t|> 1 \, ,\,  -\half < {\rm Re} \t < \half  \, , \, {\rm Im} \t > 0 \, .
}
The modular group $SL(2, \IZ) / Z_2$  for the torus is  the group of
disconnected diffeomorphisms.
The co-ordinate   transformations
$(\s_1 \, , \s_2) \rightarrow (d \s_1 + b\s_2\, , c\s_2 + a\s_1 )$
with $a, b, c, d$ integers and $ad-bc =1$, transform  the metric  into a
conformally inequivalent metric parametrized by a new modular
parameter
\eqn\modtwo{\t \rightarrow \frac{a\t +b}{c\t +d} \,  , \qquad \quad
\pmatrix{
a & b \cr
c & d \cr
} \epsilon\, SL(2, \IZ) .
}
We have to divide the $SL(2,  \IZ )$ by $Z_2$ because the elements
$\{ \1 , -\1\} $ leave
$\t$ unchanged.
In order that the theory does not suffer from global diffeomorphism anomaly,
it is necessary that the integrand in the amplitude \torusamp\  be  invariant
under the action of the modular group.
The modular group
is generated by the elements
$T: \t \rightarrow \t +1 $ and $S: \t \rightarrow  -\frac{1} {\t}$ .
Under these transformations the theta functions transform as
\eqn\modular{\eqalign{
T: \quad & \charbig{a}{b} (\t)\, \rightarrow  \, e^{-\pi i a^2 -\pi i a }
\charbig{a}{a+b+\half} (\t)\cr
S: \quad & \charbig{a}{b}  (\t)\, \rightarrow  \,  (-i\t )^\half e^{ 2 \pi i a
b}  \charbig{-b}{a} (\t)\cr
} \ . }
Moreover,
\eqn\property{\eqalign{
\char{a + m}{b +n }(\t)&= e^{2\pi i n a}\char{a}{b} (\t)\cr
\char{-a}{-b}(\t)&=\char{a}{b} (\t)\cr
} \ .}
Using these properties it is easy to check that  \torusamp\  is  modular
invariant,
and the modular integration can be restricted to
the fundamental domain \funda\ .

Let us now move on to the superstring.
For simplicity we consider the nonchiral type ${\rm II}A$ superstring
moving on $M_8 \times K_N$.  It will be easiest to use  the
Green-Schwarz formalism.
We fix the light-cone gauge by using
two of the directions in $M_8$ and obtain the remaining theory
as an orbifold. Before modding out by the orbifold group,
we have the supersymmetric sigma model in flat space
\eqn\sigmamodel{
{\cal L} = - \frac{1}{\pi} \partial_{+} X^i  \partial_{-} X^i
 -\frac{i }{\pi}S^{a} \partial_{+} S^a
 -\frac{i }{\pi} \Stil^{\ad} \partial_{-} \Stil^\ad \ ,
}
where the coordinates $X^i$  transforms as a vector
${\bf 8}_V$ of $SO(8)$ and $S^a $ and $\Stil^\ad$ transform
as spinors  of $SO(8)$, ${\bf 8}_L$  and ${\bf 8}_R$ respectively.
The orbifold group $Z_N$ is a subgroup of  planar rotations, so
we shall  use the decomposition $SO(8) \rightarrow SO(6) \times SO(2)$, or
equivalently  $SO(8) \rightarrow SU(4) \times U(1)$.
The vector and the spinor representations then decompose
as follows
\eqn\decompose{\eqalign{
{\bf 8}_V &\rightarrow {\bf 6} (0) + {\bf 1} (1)+ {\bf 1} (-1)\cr
{\bf 8}_L &\rightarrow {\bf 4} (\half) + {\bf {\bar 4}} (-\half)\cr
{\bf 8}_R &\rightarrow {\bf 4} (- \half) + {\bf {\bar 4}} (\half)\cr
} \ .}
Here the numbers in the parentheses are the  $U(1)$
charges.
We have one  boson $X$ with charge $1$,
eight  fermions $ S^m , \Stil^m$ with charge
$\half$,  and their complex conjugates. The index $m$  transforms
in the  ${\bf 4}$ of $ SU(4)$.
The one-loop vacuum amplitude
for the superstring  is quite similar  to \torusamp\ ,
\eqn\torusamp{
A_1 (N) \sim {\rm Area} \int_{\CM_1}
           {\WPM}   \quad  | \eta^2 (\t  ) \etab^2 (\tb ) \imt | ^{-3}  \quad
\frac{1}{N}\, \CZ (N) \ .
}
The area here  refers to the volume $ L^8$ of the transverse dimensions and
the first factor in the integrand comes from  the six real boson that
are neutral under  the $Z_N$ rotations.
Before discussing  the orbifold partition function $\CZ (N)$,  we should
point out  an important subtlety for the superstring that
has to do with the fact that fermions have half integer spin.
A rotation through $2\pi$ does not bring a spacetime fermion back to
itself;  which means we really have to embed our $Z_N$  not into $SO(2)$ but
into
a double cover of $SO(2)$.
As a result, it will turn out that we must distinguish between
even and odd $ N$.
Let us first consider the odd $ N$
theories.
In this case we have a sum similar to the bosonic case
\eqn\partition{
\CZ  (N_{odd}) =   \sum_{k, l = 1} ^{N}\CZ_{k,l} (N) \, .
}
Each  term $Z_{k,l}$ is a partition function
for the fields $ X, S^m , \Stil^m$ and their complex conjugates with
twisted boundary conditions
\eqn\odd{\eqalign{
S^{m}(\s_1 +1 \, , \, \s_2)  = e^{\frac{2\pi i k } {N} } \, S^{m}(\s_1 \, , \,
\s_2)\, ,
&\quad \Stil^{m}(\s_1 +1 \, , \, \s_2)  = e^{-\frac{2\pi i k } {N} } \,
S^{m}(\s_1 \, , \, \s_2)\, ,\cr
X(\s_1 +1 \, , \, \s_2)  &= e^{\frac{4\pi i k } {N} } \, X (\s_1 \, , \,
\s_2)\, \quad
\cr
} \ ,}
and similarly in the $\s_2$ direction.
Altogether, we have four fermionic and one bosonic determinants.
Using the formula  \detab\ for the determinants we obtain,
\eqn\oddpartition{
\CZ (N_{odd} ) =  \sum_{k, l = 1}^{N}
  \,  \Bigg|
\frac{  \vartheta ^4 \bigg[ {{ {\frac{k} {\sN} +\half}  \atop {\frac{l} {\sN}
+\half } }\bigg] }  }
{\eta^3 \vartheta \bigg[{ {  {\frac{2k} {\sN} +\half}  \atop  {\frac{2l}
{\sN}+\half } } \bigg] }  }
\Bigg| ^{2} \, .
}
We can repeat the analysis for even $N$ with minor modification
and obtain
\eqn\evenpartition{
\CZ (N_{even} ) =  \frac{1}{4} \sum_{k, l = 1}^{2N}
  \,  \Bigg|
\frac{  \vartheta ^4 \bigg[ {{ {\frac{k} {2\sN} +\half}
\atop {\frac{l} {2\sN}
+\half } }\bigg] }  }
{\eta^3 \vartheta \bigg[{ {  {\frac{k} {\sN} +\half}  \atop
{\frac{l} {\sN}+\half} } \bigg] }  }
\Bigg| ^{2} \, .
}
This formula would look identical  to  \oddpartition\
in terms of new variable $  N^\prime  = 2 N$,
however,  $N$ is chosen so that
$ \d = 2\pi (1-{1\over N })$.
It is straightforward to check for modular invariance.

There is one more modular invariant combination
for odd $N$ given by,
\eqn\newpartition{
{\hat\CZ} (N_{odd} ) =  \frac{1}{4} \sum_{k, l = 1}^{2N}
  \,  \Bigg|
\frac{  \vartheta ^4 \bigg[ {{ {\frac{k} {2\sN} +\half}  \atop {\frac{l} {2\sN}
+\half } }\bigg] }  }
{\eta^3 \vartheta \bigg[{ {  {\frac{k} {\sN} +\half}  \atop
{\frac{l} {\sN}+\half} } \bigg] }  }
\Bigg| ^{2} \, .
}
It  may seem a little disconcerting  that there are more
than one ways of  constructing the orbifold.  After all, if we wish to
use this construction for computing the entropy, we would like
to get a unique answer for each theory.
Fortunately, there is a  good explanation for this non-uniqueness.
The Green-Schwarz superstring has a $Z_2$ symmetry,
$(-1)^F$ where $F$ is the spacetime fermion number.
This symmetry is
obviously a subgroup of the double cover of $SO(2)$,
$(-1)^F= e^{2\pi i J_{12}}$.
The orbifold with respect to this $Z_2$  changes the
spectrum drastically. In the untwisted sector the projection
onto $Z_2$ invariant states
removes all fermions. The twisted sector adds more particles including
a tachyon.  Moreover, the number of bosonic zero modes
is the same in the twisted sectors because the bosonic coordinates
are neutral under this $Z_2$. This means that
the states in the twisted sector move over all space and are not
restricted to the tip of the cone.
We should really regard this theory
as a different theory (vacuum).
The orbifolds in \newpartition\ should then be regarded as the orbifolds not
of flat space but of this
different underlying theory.
It is to be exptected that the entropy of black holes
would be  different in these two cases  because, after all,
the patricle spectrum of the two theories is completely  different.

As an aside,  we note that in all these orbifold models,
supersymmetry is completely broken.
This is not surprising. In order to have an unbroken supersymmetry,
we must have a covariantly constant spinor on the cone, which means
that the cone must have $SU(1)$ holonomy. But $SU(1)$ holonomy is
no holonomy at all,  and the only  manifold with this holonomy is the plane.
As a result,  there are no unbroken supersymmetries on a cone.
Even though supersymmetry
is completely broken,  in some cases drastically, the
equivalence between the Green-Schwarz string and the Neveu-Schwarz-Ramond
string still continues to hold.
Let us recall the  following Riemann  theta identity:
\def\charbig#1#2{\vartheta \bigg[ {#1 \atop #2} \bigg] }
\eqn\riemann{
2 \prod_{i=1}^4 \charbig{\half}{\half} (x_i |\t)
 = \prod_{i=1}^4  \charbig{0}{0} (y_i |\t)
 - \prod_{i=1}^4  \charbig{0}{\half} (y_i |\t)
-\prod_{i=1}^4  \charbig{\half}{0} (y_i |\t)
+\prod_{i=1}^4  \charbig{\half}{\half}(y_i |\t)
}
where $y_1 = \half (x_1 + x_2 + x_3 + x_4) \, ,\,
y_2 = \half (x_1 - x_2 -  x_3 + x_4) \, ,\,
y_3 = \half (x_1 + x_2 - x_3 - x_4) \, {\rm and} \,
y_4 = \half (x_1 - x_2 + x_3 - x_4) \, $
and
$\char{a}{b} (z|\t ) = e^{2\pi i a ( z +b) } q^{a^2 \over 2} \vartheta (z+ a\t
+ b| \t )$.
With the use of this identity we can write each term in
the sum   for $\CZ (N)$ as the modulus-squared of sum of four terms.
These four terms correspond to the four spin structures on the left and the
right of the NSR superstring.
The simplest example is the ${\hat Z}_1$ orbifold above.
With $N=1$ in \newpartition\  we get,
\eqn\newpartition{
{\hat\CZ} (1 ) =  \frac { \big| {  \vartheta ^4 \big[ { 0  \atop 0  }\big] }
\big| ^{2}
+\big| {  \vartheta ^4 \big[ { 0  \atop \half  }\big] }  \big| ^{2}
+\big| {  \vartheta ^4 \big[ { \half  \atop 0  }\big] }  \big| ^{2}
+\big| {  \vartheta ^4 \big[ { \half  \atop \half  }\big] }  \big| ^{2}
}
{4 { \big|\eta^3
\vartheta \big[{    \half \atop  \half  } \big] } \big|^2  }\, \, .
}
Thus, in this case,  we see the equivalence of the
two formalisms  simply by reinterpreting the orbifold sum  in the
GS formalism as the sum over spin structures in the
NSR formalism with a modular invariant combination
diagonal in spin structures.   The corresponding NSR string  has been
discussed in \seibwitt .  It has
a very different projection than the usual GSO projection and
there are no fermions in the spectrum.
Remarkably, the GS and the NSR formalism are equivalent
even after this rather extreme breaking of supersymmetry.
This points to a deep connection between the two formalisms
that goes beyond supersymmetry.

The spectrum of  the $Z_N$ orbifold in the sector twisted by $\eta
=\frac{k}{N}$
is easily obtained.
We shall use the light-cone gauge and
describe  only the low lying states assuming that $\eta$ is small.
Let us first consider the GS formalism.
The ground state has energy $-\frac{\eta}{2}$ and is  tachyonic.
In the right moving sector, we have the
oscillator modes
$S^m_{n-\frac{\eta}{2} } \, , S^\mbar_{n+\frac{\eta}{2} } \, , \a_{n-\eta} \, ,
\, {\rm and }\, \abar_{n+\eta}$.
Acting on the vacuum with various powers of
$S^m_{-\frac{\eta}{2}}$,  we generate a sixteen
dimensional representation  quite similar to the  gauge supermultiplet
in flat space.  It  decomposes
in terms of $SU(4)$ representations  ${\bf 1}, {\bf 4}, {\bf 6}, {\bf {\bar
4}},{\bf 1}$
with masses $-\frac{\eta}{2} , 0 ,\frac{\eta}{2}, \eta , \frac{3 \eta}{2}$
respectively.
The representations ${\bf 4}$ and ${\bf {\bar 4}}$
are fermions and the remaining states are
bosons.
We get  a similar representation on the left and
the low lying spectrum is
the tensor product of the two,  keeping only
the  $Z_N$ invariant states. In addition,  there are more states that are
obtained by acting with   various powers of $\a_{-\eta}$ on these states.
In the NSR formalism the analysis is somewhat different.
The eight worldsheet fermions $\psi^i$ transform as the vector of $SO(8)$
exactly like the bosons .
As a result, only two get twisted and the remaining six are untwisted.
In the NS sector,
the ground state energy is $-\half +\frac{\eta}{2}$.
The low energy fermion creation operators are
 $\psi_{-\half - \eta  } $  and $ {\bar \psi}_{-\half + \eta }$  coming from
the two twisted fermions,  and six
 $\psi^i_{-\half } $ coming from the untwisted fermions.
The ground state gets projected out by the GSO projection.
At the next level,  we have one of
 the creation operators above acting on the vacuum.
This gives one  state with energy $-\frac{\eta}{2}$,   six states with energy
$\frac{\eta}{2}$
and one state with
energy $\frac{3 \eta}{2}$.
All of these are spacetime bosons.  As  usual,
the spacetime fermions come from the Ramond sector.
The ground state energy is zero in the Ramond sector.
The six untwisted fermions have zero modes that
form the Clifford algebrea of $SO(6)$ and have
an eight-dimensional spinor representation,
that splits as ${\bf 4} $ and ${\bf {\bar 4}}$ . The ${\bf {\bar 4}}$
gets projected out by the GSO projection and  we are left with four fermions
with mass $0$ in ${\bf 4} $.  The next excited state is obtained
by acting on the vacuum with  a creation operator  with energy $\eta$
coming from the twisted fermions.  The GSO projection removes the
 ${\bf 4} $ at this level  and we obtain four fermions in ${\bf {\bar 4}}$
with energy $ \eta$.
The bosonic oscillators in the NSR string are  the same as in
the GS string,  so  the low lying spectrum matches
exactly with the one obtained from  the GS formalism.

 \newsec{  Black Hole Entropy in String Theory}

The entropy is given by,
\eqn\entropydef{
S = -\b \frac{\partial (\log Z)}{\partial \b} +\log Z.}
In order to compute this we need to vary the Rindler
temperature away from $2\pi$ which corresponds to flat
space. This introduces a conical defect with the deficit
angle $\delta$ which is related to the inverse temperature by
$\b = 2\pi (1 - \frac{\d}{2\pi} ) =  {2\pi  \over N } $.
Treating $N$ as a continuous variable,  we see that
$S = \frac{d (N \, \log {Z} ) }{dN} |_{N=1}$ .
In string perturbation theory at $G$ loop,
the spacetime partition function $Z_{G} $  and the
worldsheet vacuum amplitude $A_{G}$ are related by
$\log{ Z_{G}} = A_{G}$ \tseytlin . This gives us the desired formula for
the entropy at $G$ loop:
\eqn\entropy{
S_{G}  \, = \, \frac{d (N \, A_{G} ) }{dN} |_{N=1} \, .
}
Substituting \torusamp\  into \entropy\ we obtain the final
expression for the entropy in the bosonic string at one loop:
\eqn\entropybose{
S_1  \sim {\rm Area} \int_{\CM_1}
           {\WPM}   \quad  | \eta^2 (\t  ) \etab^2 (\tb ) \imt | ^{-11}  \,
\CZ ^{\prime}(1)  \quad .
}
Similarly for the type IIA supersting we obtain
\eqn\entropysuper{
S_1  \sim {\rm Area} \int_{\CM_1}
           {\WPM}   \quad  | \eta^2 (\t  ) \etab^2 (\tb ) \imt | ^{-3}  \,
\CZ ^{\prime}(1)  \quad .
}
Here $ \CZ ^{\prime}(1)$  for each theory is  the first derivative
of the orbifold partition function evaluated at $N=1$.
It is quite satisfying that the entropy comes out proportional to the area.
Before making this assertion, we must clarify  one point  that we have
so far glossed over.
In the orbifold sum
\partition\  the term  $\CZ_{0, 0}$ is divergent.
This is because the theta function vanishes
due to the contribution of  the  bosonic zero modes
to the corresonding determinant.
If  we treat  the determinant  carefully,  the zero modes
would turn the area in \torusamp\ into volume
for this particular term in the sum.
Modular invariance would still hold because $\CZ_{0,0}$ does not
mix with other terms under modular transformations and is invariant
by itself. More importantly,  it  is independent of $N$ and
as we see from \entropybose\ ,  it does not contribute to the entropy.
Thus, the entropy will be proportional
to the area and not the volume.

Before proceeding further let us see what we expect to find.
The Schwarzschild observer near the horizon sees a hot thermal
bath. We can thus view the orbifolds
as describing the Euclidean path integral for strings
in a thermal ensemble at Rindler temperature ${N\over 2\pi}$.
The proper temperature is position dependent and diverges
at the tip of the cone. As a result, we expect the strings
to undergo a Hagedorn phase transition well known in string
theory. Correspondingly, we expcect to find an infrared instability
in the form of tachyons.
The tachyons that we find in the twisted sector are closely
related to tachyons coming from the winding modes of the string
around the Euclidean time direction \satkog\ which signal
the Hagedorn transition in strings at finite temperature.
After the phase transition, a tachyon condensate will be formed.
This can explain the tree level contribution to the entropy
as coming from the latent heat of this phase transition \aticwitt .
Furthermore this condensate will be confined very close to the horizon
and spread only in the transverse directions. It raises
the exciting possibility that we can understand
the dependence of the Bekenstein-Hawking
entropy on area as well as the coupling constant \inprep .
Unfortunately, at the moment, the Hagedorn transition is not very
well understood. Moreover, once we include interactions
we also have to worry about the Jeans instability \bowgidd .
It is not clear how to properly take these effects into
account. However, we are really interested only in the entropy
at a special value of the temperature and we hope that at least some
of the features of the entropy will be accesible without having
to understand all the consequences of the Hagedorn transition.
For example, it would be  nice to see if the entropy can be
rendered finite in the infrared by adding a tree level
contribution.

In order to complete this computation  we need  an analytic expression
for the sum $\CZ (N)$ so that we can take its derivative. Several comments
are in order  here. First, all terms in the sum are analytic
functions
of $N$ so we can expect  that the sum will also be analytic.
Second,  modular invariance of the vacuum amplitude
holds only for integer $N$. In fact,  for non-integer $N$,
the sum may not have
any interpretation  as a partition function
of some string theory. However,  for our entropy computation
we do not really require that the theory be
well defined for arbitrary $N$.
All  that is needed is that
the first derivative of the vacuum amplitude
at $N=1$ be well-defined and modular invariant.
This certainly seems possible, especially if we think of the entropy
as the counting of states of a given finite theory in flat space as seen by
the Rindler observer.
Finally,  even if  we do  perform the
sum,  there is a high degree of non-uniquness because
we can always add to the sum any function that vanishes
for integer $N$  \eg  \ $ \sin (\pi N)$.  We can fix this non-uniqueness
if  we know that  the partition function does not
have an essential singularity at $N=\infty$.
This requires some physical input about the theory.
We can guess the correct ananlytic continuation
by comparing our answers with strings in a thermal ensemble
discussed in \inprep .

It is interesting to take the
large N limit of our formula \partition\ . Putting $z = \frac{k}{N} \t +
\frac{l}{N} $ in \partition\ and taking $N$ to infinity,
we see that
\eqn\largeN{
\CZ (N) |_{N\rightarrow \infty} =
\frac{N^2}{\imt}  \int d^2 z  \ {\rm exp }(- \frac{\pi (z -\zbar)^2}{2 \imt} )\
 \big| \frac{ \eta (\t )} { \vartheta_{11} (z |\t )} \big|^{2}
}
where $\vartheta_{11}$ is the theta function with half characteristics.
The integration is over a torus with modular parameter $\t$
which is parametrized with a flat metric as
a parallelogram with corners at $0$, $1$, $\t$ and $\t +1$.
The integrand is doubly periodic over this region and  is
a well defined function on the torus.
This expression is clearly analytic in $N$ and moreover is modular invariant
even for non-integer $N$.
Encouragingly,  $\CZ (N)$ does not have
an essential singularity but only a  second order pole at infinite $N$.
Another interesting feature of \largeN\ is that  the integral is
logarithmically divergent.  The theta function has a zero at $z=0$,
${ \vartheta_{11} (z |\t )} \sim 2\pi   z \eta^3  (\t )  $.
As a result the leading contribution to \largeN\  is
\eqn\largeNN{
\CZ (N) |_{N\rightarrow \infty}  \sim N^2 \log N \ | \eta^2 (\t  ) \etab^2 (\tb
) \imt | ^{-1} \ .
}
This expression is strikingly similar to the partition function in $R_2 $  $
(N=1)$ after
properly taking into account the bosonic zero modes
\eqn\plane{
\CZ (N) |_{N = 1}  \sim L^2\
| \eta^2 (\t  ) \etab^2 (\tb ) \imt | ^{-1} \ .
}
We regard this as evidence for some kind of duality similar
to the ${\scriptstyle R}\rightarrow {1\over R} $ duality.
As $N$ becomes large,
the space is becoming smaller
and one might think that
the number of states is also becoming smaller.
However, more and more twisted states come down and
become almost massless as we take $N$ to infinity.
These states combine to give a partition
function very similar to the partition function of the original theory.
So in
some sense we still have as many states as we started with.

\newsec{Discussion}

We have seen that string theory offers a suitable framework for
testing the conjecture by `t Hooft and Susskind
that the Bekenstein-Hawking
entropy should be understood in terms of the entropy of
fluctuations near the horizon. With this objective in mind,
we have described the construction of string propagation on a cone.
We have obtained an  expression for the entropy that is proportional
to the area of the event horizon.
In order to complete this computation we need to perform the finite
sum for $\CZ (N)$ which is currently under investigation \inprep .
The presence of tachyons in these models also deserves attention.
Tachyons signify a vacuum
instability and it is very important to understand their
role in string theory. For example,
it has long been thought that the bosonic string represents
a metastable point in the space of vacua and in the proper non-perturbative
formulation we would see the theory  relax into one of the stable ground
states.
This proposition is of course too difficult to test in  flat place because
the usual tachyon moves over all  space and
we do not even know any candidates for a nearby ground states.
In our case,  we have a whole family of theories which
can have a large number of tachyons localized to the tip of the cone.
If a tachyon condensate is formed, it will most likely change
the value of the deficit angle \ie the value of  $N$.
For large N,  some of
the tachyons are nearly massless and it may be possible to understand
their condensation as a perturbation of the $K_N$ conformal field
theory with a nearly marginal operator.

The existence of tachyons in the
twisted sectors along with the results
of \inprep\ indicates that there will be a Hagedorn transition
close to horizon.
It seems possible to understand the
dependence of the black hole entropy on the area of the event
horizon and also the coupling constant if a
condensate is formed. If the entropy comes from the
fundamental degrees of freedom of string theory beyond the
Hagedorn transition, then quite possibly, it is also independent
of the low energy particle spectrum.
If this conjecture turns out to be correct, we may be able to learn
something about the Hagedorn transition from its relation to the
black hole entropy and vice versa.

\bigskip
\leftline{ \secfont Acknowledgements}
\bigskip
I would like to thank Samir Mathur, Stephen Shenker and especially Cumrun
Vafa for useful  discussions, and the organizers of the
`Second International Colloquium on
Modern Quantum Field Theory' at the Tata Institute for Fundamental Research,
for their hospitality.
This research is supported in part by the Packard Foundation, NSF
grants PHY 89-57162 and PHY-92-18167 .

\listrefs
\end